\documentclass[amsart]{article}
\usepackage{authblk}
\usepackage{graphics}
\usepackage{caption}
\usepackage{mathtools}
\usepackage{amsmath}
\usepackage{mathrsfs}
\usepackage{graphicx}
\usepackage[flushleft]{threeparttable}
\usepackage[utf8]{inputenc}
\usepackage{wrapfig}
\usepackage{epstopdf}
\usepackage{enumitem}
\usepackage{lipsum}
\usepackage{placeins}
\usepackage[margin=1in]{geometry}
\usepackage{abstract}

\title{Thermodynamic relations among isotropic material properties in conditions of plane shear stress}
\date{}
\author[1,2,*]{Amilcare Porporato}
\author[1]{Salvatore Calabrese}
\author[3]{Tomasz Hueckel}

\affil[1]{Department of Civil and Environmental Engineering, Princeton University, Princeton, NJ, USA.}
\affil[2]{Princeton Environmental Institute, Princeton University, Princeton, NJ, USA.}
\affil[3]{Department of Civil and Environmental Engineering, Duke University, Durham, NC, USA.}
\affil[*]{Corresponding author: Amilcare Porporato, aporpora@princeton.edu}

\begin{document}
	\maketitle
	
\section*{Abstract}

We present new general relationships among the material properties of an isotropic material kept in homogeneous stress conditions with hydrostatic pressure and plane shear. The derivation is not limited to the proximity of the zero shear-stress and -strain condition, which allows us to identify the relationship between adiabatic and isothermal shear compliances (inverse of the moduli of rigidity) along with new links, among others, between isobaric and isochoric shear thermal expansion coefficients, heat capacities at constant stress and constant shear strain. Such relationships are important for a variety of applications, including characterization of nanomaterials as well as identification of properties related to earthquakes precursors and complex media (e.g., soil) behaviors. The results may be useful to investigate the behavior of materials during phase transitions involving shear or in non-homogeneous conditions within a local thermodynamic equilibrium framework.
	
\section{Introduction}

Since Gibbs' fundamental contribution in 1876 \cite{gibbs1928collected}, the thermodynamic theory of solids under different stress conditions has remained an active field of inquiry, with a recent intensification spurred by interest in amorphous states and glass transition, high pressure physics and the development of artificial materials \cite{poirier2000introduction,debenedetti2001supercooled,dyre2006colloquium,charbonneau2017glass,xu2018predicting,burns201877}. In contrast, continuum mechanics and thermoelasticity have focused more on finite deformations and field theories, traditionally shifting away from homogeneous thermodynamics \cite{guzzetta2013relating,berezovski2017internal} and the related 
Gibbs equation \cite{truesdell2004non}, in spite of the fact the these concepts are often clearer, at least for infinitesimal transformations and in uniform conditions \cite{berezovski2017internal}.

One of the great utilities of equilibrium thermodynamics lies in its theoretical structure, which provides fundamental links among material properties (such as heat capacities, compressibilities and thermal expansion) through the Hessian matrix of the (generalized) Gibbs free energy, while ensuring feasible reversible transformations (the so-called stability conditions \cite{callen2006thermodynamics}). The relationships between material properties in hydrostatic conditions are well known since Clapeyron \cite{brillouin1940influence,menikoff1989riemann,callen2006thermodynamics} and much research has been carried out, in such conditions, on linking these properties to the equations of state and the fundamental equation, especially at high pressures of interest to geophysics and astrophysics \cite{poirier2000introduction,ponkratz2004equations,heuze2012general,holzapfel2018coherent}.

For solids in non hydrostatic conditions the number of independent material properties quickly grows. The relevant ones, including latent heats, have been defined and discussed since Kelvin and Gibbs \cite{fung2017foundations}. The complete relationships among them and a full discussion of their meaning have been missing, however, due also in part to the complex stress patterns of crystals and anisotropic solids \cite{gibbs1928collected,born1939thermodynamics,landau1965theory,fung2017foundations}. Only very recently, Burns \cite{burns201877} has provided a list of these relationships obtained using the Jacobian algebra of thermodynamic transformations.

Here we focus on the case of plane shear, which is intermediate between the hydrostatic and the fully anisotropic one. While it represents a highly idealized state compared to the heterogenous and anisotropic stress configurations typical of real life conditions, it remains an important benchmark for averaged properties of poli-crystals and amorphous materials. The case of homogeneous and isotropic but non hydrostatic stress is dealt with in classic texts \cite{landau1965theory,poirier2000introduction,fung2017foundations}. Their thermodynamic analysis, however, is essentially limited to the case of small deformations around the state of zero shear stress and deformation. This is a very important but special case, where there is equality of the isothermal and isentropic shear compliance (or their inverse, i.e., the moduli of rigidity $\mu_T$ and $\mu_s$), as pointed out by Brillouin \cite{brillouin1940influence}, and mentioned also in other work \cite{landau1965theory,poirier2000introduction,dyre2006colloquium}. With the exception of Burns \cite{burns201877,burns2018elastic}, this condition, however, is treated as one of general validity. 

Burns has also drawn attention to the general relationships among material properties in crystals \cite{burns201877} and in conditions of plane shear \cite{burns2018elastic}, finding new explicit relationships and discussing special cases. Our contribution here represents an extension to his second work \cite{burns2018elastic}. In particular, we relax the condition of zero hydrostatic pressure and, most importantly, we do not limit the discussion to transformations around the zero shear and strain point and include a coefficient of dilatancy \cite{reynolds1885lvii}, expressing the coupling between pressure and shear stress. We provide a complete discussion of quasi-static transformations and the full spectrum of relationships among material properties, which show, \textit{inter alia}, how the equality between the isothermal and isentropic rigidity moduli requires constant shear stress.

\section{The Gibbs equation and generalized free energy}
Consider an isotropic material in conditions of homogeneous stress, given by a hydrostatic pressure, $p$, and a plane shear stress, $\tau$, which derive from decomposing the stress tensor $\boldsymbol{\sigma}$ \cite{landau1965theory} as
\begin{equation}
	\label{eq:stresscond}
\boldsymbol{\sigma} = 
\left(\begin{matrix}
 p & 0 & 0 \\
 0 & p & 0 \\
 0 & 0 & p
\end{matrix}\right)+
\left( \begin{matrix}
0 & \tau &0 \\
\tau & 0 & 0 \\
0 & 0& 0
\end{matrix}\right).
\end{equation}

The isotropic material in the stress state (\ref{eq:stresscond}) represents a thermodynamic system, whose internal energy $U$ is a state function related to the other state variables defining the thermodynamic state through the fundamental equation 
\begin{equation}
\label{eq:fundequ}
U=U(S,V,N,\gamma),
\end{equation}
where $S$, $V$, and $N$ are the entropy, the volume and the number of moles, respectively, and $\gamma$ is the shear strain (as a reference, $\gamma=0$ in hydrostatic conditions). For reversible transformations, the total differential of (\ref{eq:fundequ}) is exact and reads
\begin{equation}
\label{eq:gibbsshear1}
dU=TdS-pdV+\mu dN +\tau V d\gamma,
\end{equation}
where the temperature, $T$, pressure, $p$, chemical potential, $\mu$, and shear stress, $\tau$, are partial derivatives of the internal energy,
\begin{equation}
T=\left.\frac{\partial U}{\partial S}\right|_{V,N,\gamma}, \quad \quad p=-\left.\frac{\partial U}{\partial V}\right|_{S,N,\gamma}, \quad \mu=\left.\frac{\partial U}{\partial N}\right|_{S,V,\gamma} \quad \text{and} \quad \tau=\frac{1}{V}\left.\frac{\partial U}{\partial \gamma}\right|_{S,V,N}.
\end{equation}
Equation (\ref{eq:gibbsshear1}) is the Gibbs equation, governing the conservation of energy during infinitesimal reversible transformations around a generic equilibrium state \cite{li1978physical,landau1965theory}. Specifically, the change, $dU$, in internal energy can be due to the term $TdS$, representing the heat exchanged, or to the work terms $pdV$, $\mu dN$ and $\tau V d\gamma$ that are due to expansion/compression, change in the number of moles, and change in shear angle, respectively. 

It is more convenient here to express (\ref{eq:gibbsshear1}) for constant mass, i.e., $dN=0$, and per unit volume by dividing it by $V$,
\begin{equation}
\label{eq:gibbsshear}
du=Tds-pd\epsilon+\tau d\gamma.
\end{equation}
where $du=dU/V$, $ds=dS/V$, $d\epsilon=dV/V$ is the incremental volumetric strain, where $\epsilon=\ln(V/V_0)$ is the logarithmic strain \cite{hencky1928uber,fitzgerald1980tensorial} with $V_0$ a reference volume. 

The material properties are derived from the corresponding extended Gibbs free energy per unit volume, which is a function of $T$, $p$, and $\tau$, and reads
\begin{equation}
g(T,p,\tau)=u-Ts+p\epsilon -\tau \gamma.
\end{equation}
Using (\ref{eq:gibbsshear}), its differential becomes
\begin{equation}
dg=-sdT+\epsilon dp -\gamma d\tau,
\end{equation}
where
\begin{equation}
s=-\left.\frac{\partial g}{\partial T}\right|_{p,\tau}, \quad \quad \epsilon=\left.\frac{\partial g}{\partial p}\right|_{T,\tau}, \quad \text{and} \quad \gamma=-\left.\frac{\partial g}{\partial \tau}\right|_{T,p}.
\end{equation}

From the extended Gibbs free energy, $g$, the material properties are found as the entries of its Hessian matrix, namely they are given by the second derivatives of $g$. From the diagonal components of the Hessian matrix, we have the heat capacity at constant pressure and shear stress
\begin{equation}
c_{p,\tau}=-T\left.\frac{\partial^2 g}{\partial T^2}\right|_{p,\tau}=T\left.\frac{\partial s}{\partial T}\right|_{p,\tau},
\end{equation}
the isothermal, isoshear compressibility
\begin{equation}
k_{T,\tau}=\left.\frac{\partial^2 g}{\partial p^2}\right|_{T,\tau}=-\left.\frac{\partial \epsilon}{\partial p}\right|_{T,\tau},
\end{equation}
and the isothermal, isobaric shear compliance
\begin{equation}
S_{T,p}=-\left.\frac{\partial^2 g}{\partial \tau^2}\right|_{T,p}=\left.\frac{\partial \gamma}{\partial \tau}\right|_{T,p}.
\end{equation}

The properties describing the coupling between thermal, pressure and shear transformations are defined by the cross derivatives of the Gibbs free energy and are the off-diagonal components of the Hessian: the coefficient of thermal expansion at constant pressure and shear stress
\begin{equation}
\alpha_{\tau}= \left.\frac{\partial^2 g}{\partial p \partial T}\right|_{\tau}=\left.\frac{\partial \epsilon}{\partial T}\right|_{p,\tau}=-\left.\frac{\partial s}{\partial p}\right|_{T,\tau},
\end{equation}
the coefficient of thermal shear deformation at constant pressure and shear stress, which is also a coefficient of entropy change due to shear stress at constant temperature and pressure,
\begin{equation}
\beta_{p}=-\left.\frac{\partial^2 g}{\partial \tau \partial T}\right|_{p}=\left.\frac{\partial \gamma}{\partial T}\right|_{p,\tau}=\left.\frac{\partial s}{\partial \tau}\right|_{T,p},
\end{equation}
and the coefficient of isothermal shear deformation due to pressure change at constant shear stress, which is also the coefficient of dilatancy, namely the expansion due to change in shear stress at constant temperature and pressure,
\begin{equation}
\eta_{T}=\left.\frac{\partial^2 g}{\partial \tau \partial p}\right|_{T}=-\left.\frac{\partial \gamma}{\partial p}\right|_{T,\tau}=\left.\frac{\partial \epsilon}{\partial \tau}\right|_{T,p}.
\end{equation}
\begin{table}
	\label{tab:1}
		\centering
			\caption{Material properties as derived from the Gibbs free energy. The first row contains the extensive variable to differentiate, while the first column contains the operators.}
		\begin{tabular}{l|c|c|l}
			  & $s$ & $\epsilon$  &$\gamma$  \\
			\hline
			  $\frac{\partial  }{\partial T}$ & $\frac{c_{p,\tau}}{T}$ & $\alpha_{\tau}$ & $\beta_p$ \\
			$\frac{\partial }{\partial p}$ & -$\alpha_{\tau}$& -$k_{t,\tau}$ & -$\eta_T$ \\
			 $\frac{\partial  }{\partial \tau}$  &$\beta_p$ &  $\eta_T$& $S_{T,p}$ \\
					
		\end{tabular}
\\	{\tiny }
\end{table}
Being considered as an earthquake precursor \cite{rowe1962stress,brace1966dilatancy,nur1972dilatancy,barnes1989shear}, this latter coefficient has a paramount practical importance, since its discovery by Reynolds \cite{reynolds1885lvii}, in a variety of granular materials, solid suspensions, soils and rocks, as well as in a 2-D Schneebeli analogue material (a bunch of pencil-like rods)\cite{schneebeli1956analogie,sibille2007numerical}.

The Hessian matrix of the Gibbs free energy thus provides the six independent material properties necessary to describe the thermodynamics behavior (see Table 1). A total of 18 material properties (6 independent and 12 dependent) exists depending on the thermodynamic constraints imposed (e.g., isothermal, adiabatic, isobaric, and so on). With subscripts identifying the variables held constant, we have four heat capacities ($c_{v,\gamma}$, $c_{v,\tau}$, $c_{p,\gamma}$, and $c_{p,\tau}$), four compressibilities ($k_{T,\gamma}$, $k_{T,\tau}$, $k_{s,\gamma}$, and $k_{s,\tau}$), four compliances ($S_{T,v}$, $S_{T,p}$, $S_{s,v}$, and $S_{s,p}$), two coefficients of thermal expansion ($\alpha_{\tau}$ and $\alpha_{\gamma}$), two coefficients of thermal shear deformation ($\beta_{\epsilon}$ and $\beta_{p}$), and two coefficients of shear deformation due to pressure or dilatancy ($\eta_T$ and $\eta_s$).

\section{Relations among the material properties}
These material properties, described above, define the curvature of the state functions, such as $u$, $s$ or $\epsilon$, in the space of their independent variables. For infinitesimal reversible transformations, they can be employed to infer the change (or total differential) of any state variable, as shown below. It is convenient to focus on the representations provided by the independent variables in (\ref{eq:fundequ}). These are the well-known entropy representation \cite{callen2006thermodynamics} (section 3.1) and the less common volume and shear angle representation (sections 3.2 and 3.3, respectively).

%Defining the curvature of the Gibbs free energy, the material properties can be employed to infer the path along the free energy surface during reversible transformations, namely the total differential of a thermodynamic variable (either extensive or intensive) can be formulated in terms of them. Also, from the Gibbs free energy other representations that can be more convenient to interpret thermodynamics processes are obtained by Legendre transformation. Below, we consider the entropy, volume, and angle of deformation representations, which, nevertheless, could be rearranged to yield the total differentials of $T$, $p$ and $\tau$.

\subsection{Entropy representation}
The equilibrium state of the material, expressed in energy representation in equation (\ref{eq:fundequ}), can analogously be described in entropy representation \cite{callen2006thermodynamics} as
\begin{equation}
s=s(u,\epsilon,\gamma).
\end{equation}
Through a change of variables, the entropy $s$ can then be written as a function of the independent variables $T$, $p$, and $\tau$, $s=s(T,p,\tau)$, such that the variation $ds$ during an infinitesimal transformation reads
\begin{equation}
ds=\left.\frac{\partial s}{\partial T}\right|_{p,\tau}dT+\left.\frac{\partial s}{\partial p}\right|_{T,\tau}dp+\left.\frac{\partial s}{\partial \tau}\right|_{T,p}d\tau.
\end{equation}
Substituting the partial derivatives with the material properties in Table 1, one obtains
\begin{equation}
ds=\frac{c_{p,\tau}}{T}dT-\alpha_{\tau}dp+\beta_{p}d\tau.
\end{equation}
Using different combinations of thermodynamic variables as independent quantities and introducing the corresponding material properties, the total differential can be written in the following forms,
% i.e., as $s=s(T,p,\gamma)=s(T,\epsilon,\tau)=s(T,\epsilon,\gamma)$. Differentiating and substituting the derivatives with the corresponding material properties, the differentials can be written accordingly as
\begin{eqnarray}
ds=\frac{c_{p,\gamma}}{T}dT-\alpha_{\gamma}dp+\frac{\beta_p}{S_{T,p}}d\gamma=\frac{c_{p,\gamma}}{T}dT-\alpha_{\gamma}dp+\frac{\lambda_{\gamma,p}}{T}d\gamma,\\
ds=\frac{c_{v,\tau}}{T}dT+\frac{\alpha_{\tau}}{k_{T,\tau}}d\epsilon+\beta_\epsilon d\tau=\frac{c_{v,\tau}}{T}dT+\frac{\lambda_{\epsilon,\tau}}{T}d\epsilon+\beta_\epsilon d\tau\\
ds=\frac{c_{v,\gamma}}{T}dT+\frac{\alpha_{\gamma}}{k_{T,\gamma}}d\epsilon+\frac{\beta_\epsilon}{S_{T,\epsilon}}d\gamma=\frac{c_{v,\gamma}}{T}dT+\frac{\lambda_{\epsilon,\gamma}}{T}d\epsilon+\frac{\lambda_{\gamma,\epsilon}}{T}d\gamma,
\end{eqnarray}
where we introduced the latent heats of volumetric expansion, $\lambda_{\epsilon,\tau}=\frac{\alpha_\tau T}{K_{T,\tau}}$, $\lambda_{\epsilon,\gamma}=\frac{\alpha_\gamma T}{K_{T,\gamma}}$, and of shear deformation, $\lambda_{\gamma,p}=\frac{\beta_p T}{S_{T,p}}$ and $\lambda_{\gamma,\epsilon}=\frac{\beta_\epsilon T}{S_{T,\epsilon}}$ \cite{fung2017foundations}.
%\begin{eqnarray}
%ds=\frac{c_{p,\gamma}}{T}dT-\alpha_{\gamma}dp+\frac{\lambda_{\gamma,p}}{T}d\gamma=\frac{c_{v,\tau}}{T}dT+\frac{\lambda_{\epsilon,\tau}}{T}d\epsilon+\beta_\epsilon d\tau=\\\frac{c_{v,\gamma}}{T}dT+\frac{\lambda_{\epsilon,\gamma}}{T}d\epsilon+\frac{\lambda_{\gamma,\epsilon}}{T}d\gamma,
%\end{eqnarray}
By equating the above expressions for the total differential of the entropy, one readily obtains interesting relationships among thermal properties, i.e., heat capacities, coefficients of thermal expansion and thermal shear deformation, and pressure and shear properties, i.e., compliances and coefficients of dilatancy.

\subsubsection{\boldmath$ds(T,p,\tau)=ds(T,p, \bf \gamma)$}
From
\begin{equation}
\frac{c_{p,\tau}}{T}dT-\alpha_{\tau}dp+\beta_{p}d\tau=\frac{c_{p,\gamma}}{T}dT-\alpha_{\gamma}dp+\frac{\beta_p}{S_{T,p}}d\gamma,
\end{equation}
dividing by $d\gamma$ at constant $\tau$, and substituting in the material properties, one obtains
\begin{equation}
\label{eq:entrtransf1}
\frac{c_{p,\tau}-c_{p,\gamma}}{\beta_p T}+\frac{(\alpha_{\tau}-\alpha_{\gamma})}{\eta_T}=\frac{\beta_p}{S_{T,p}}.
\end{equation}
The same result is obtained if one divides by $d\tau$ at constant $\gamma$. 

Since $p$ appears on both sides of $ds(T,p,\tau)=ds(T,p,\gamma)$, we could consider a further condition, $dp=0$, that is a isobaric transformation. In such a case, (\ref{eq:entrtransf1}) reduces to
\begin{equation}
c_{p,\tau}=c_{p,\gamma}+\frac{\beta_p^2 T}{S_{T,p}},
\end{equation}
and thus
\begin{equation}
\alpha_{\tau}=\alpha_{\gamma},
\end{equation}
namely at constant $p$ the coefficient of thermal expansion does not depend on whether $\tau$ or $\gamma$ is held constant. 

If $T$ is held constant, rather than $p$, one obtains
\begin{equation}
\alpha_{\tau}=\alpha_{\gamma}+\frac{\beta_p \eta_T}{S_{T,p}},
\end{equation}
and thus
\begin{equation}
c_{p,\tau}=c_{p,\gamma}.
\end{equation}

\subsubsection{\boldmath$ds(T,p,\gamma)=ds(T,\epsilon,\tau)$}

From this equality, again dividing by $d\gamma$ at constant $\tau$, one has
\begin{equation}
\label{eq:entrtransf2}
\frac{c_{v,\tau}}{\beta_\epsilon T}-\frac{c_{p,\gamma}}{\beta_{p} T}+\frac{\alpha_{\tau}-\alpha_{\gamma}}{\eta_{T}}=\frac{\beta_p}{S_{T,p}}.
\end{equation}

Proceeding as for equation (\ref{eq:entrtransf1}), along an isotherm, $dT=0$, one obtains again
\begin{equation}
\alpha_{\tau}=\alpha_{\gamma}+\frac{\beta_p \eta_{T}}{S_{T,p}}.
\end{equation}
and hence
\begin{equation}
\frac{c_{p,\gamma}}{c_{v,\tau}}=\frac{\beta_{p}}{\beta_\epsilon}.
\end{equation}

\subsubsection{\boldmath$ds(T,\epsilon,\tau)=ds(T,\epsilon,\gamma)$}
Following the same procedure, we have
\begin{equation}
\label{eq:entrtransf3}
\frac{c_{v,\tau}-c_{v,\gamma}}{\beta_\epsilon T}+\frac{1}{\eta_{T}}\left(\alpha_{\tau}-\alpha_{\gamma}\frac{k_{T,\tau}}{k_{T,\gamma}}\right)=\frac{ \beta_\epsilon}{S_{T,\epsilon}}.
\end{equation}
For $dT=0$,
\begin{equation}
\alpha_{\tau}=\alpha_{\gamma}\frac{k_{T,\tau}}{k_{T,\gamma}}+\frac{ \beta_\epsilon \eta_{T}}{S_{T,\epsilon}},
\end{equation}
and as a consequence $c_{v,\tau}=c_{v,\gamma}$.
For $d\epsilon=0$,
\begin{equation}
c_{v,\tau}=c_{v,\gamma}+\frac{ \beta_\epsilon^2 T}{S_{T,\epsilon}},
\end{equation}
and thus
\begin{equation}
\frac{\alpha_{\tau}}{\alpha_{\gamma}}=\frac{k_{T,\tau}}{k_{T,\gamma}}.
\end{equation}

\subsubsection{\boldmath$ds(T,\epsilon,\gamma)=ds(T,p,\tau)$}
Finally,  we have
\begin{equation}
\label{eq:entrtransf4}
\frac{c_{p,\tau}-c_{v,\gamma}}{ \beta_{\epsilon} T}+\frac{1}{\eta_{T}}\left(\alpha_{\tau}-\alpha_{\gamma}\frac{k_{T,\tau}}{k_{T,\gamma}}\right)=\frac{\beta_{\epsilon}}{S_{T,\gamma}},
\end{equation}
which along an isotherm, $dT=0$, reduces to
\begin{equation}
\alpha_{\tau}=\alpha_{\gamma}\frac{k_{T,\tau}}{k_{T,\gamma}}+\frac{\beta_{\epsilon} \eta_{T}}{S_{T,\gamma}},
\end{equation}
and thus $c_{p,\tau}=c_{v,\gamma}$.

\subsection{Volume representation}
The equilibrium state can also be described in the volume representation. Proceeding as for the entropy representation, depending on the choice of independent variables, the total differential of the volumetric expansion $\epsilon$ can be expressed in terms of various material properties as follows
\begin{eqnarray}
\label{eq:expdiff}
d\epsilon=\alpha_\tau dT-k_{T,\tau}dp+\eta_{T}d\tau\\
d\epsilon=\frac{\alpha_{\tau} T}{c_{p,\tau}}ds-k_{s,\tau}dp+\eta_s d\tau \\
d\epsilon=\alpha_{\gamma}dT-k_{T,\gamma}dp+\frac{\eta_T}{S_{T,p}}d\gamma\\
d\epsilon=\frac{\alpha_{\gamma} T}{c_{p,\gamma}}ds-k_{s,\gamma}dp+\frac{\eta_s}{S_{s,p}}d\gamma.
\end{eqnarray}
Similarly to the previous section, from the above differentials, relationships between pressure properties, e.g., compressibility, and thermal and shear properties can be derived.

\subsubsection{\boldmath$d\epsilon(T,p,\tau)=d\epsilon(s,p,\tau)$}
\label{sec:volmrep1}
Division by $dS$ at constant $T$ yields
\begin{equation}
\label{eq:voltransf1}
\frac{k_{T,\tau}-k_{s,\tau}}{\alpha_{\tau}}+\frac{\eta_T-\eta_s}{\beta_{p}}=\frac{\alpha_\tau T}{c_{p,\tau}}.
\end{equation}
Along an isoshear-stress, $d\tau=0$, the well-known relationship between isothermal and adiabatic compressibility is recovered \cite{callen2006thermodynamics},
\begin{equation}
\label{eq:krel1}
k_{T,\tau}=k_{s,\tau}+\frac{\alpha_\tau^2 T}{c_{p,\tau}},
\end{equation}
where we emphasize that equation (\ref{eq:krel1}) is generally obtained for hydrostatic conditions, while here it is extended to a generic condition of non-zero plane shear stress, as long as the shear stress remains constant.
From (\ref{eq:voltransf1}) and $d\tau=0$, one also gets
\begin{equation}
\eta_T=\eta_s.
\end{equation}
Along an isobar, $dp=0$,
\begin{equation}
\eta_T=\eta_s+\frac{\alpha_\tau \beta_{p} T}{c_{p,\tau}},
\end{equation}
and thus
\begin{equation}
k_{t,\tau}=k_{s,\tau}.
\end{equation}

\subsubsection{\boldmath$d\epsilon(s,p,\tau)=d\epsilon(T,p,\gamma)$}
Analogously to section \ref{sec:volmrep1}, we obtain
\begin{equation}
\label{eq:voltransf2}
\frac{k_{T,\gamma}}{\alpha_{\gamma}}-\frac{k_{s,\tau}}{\alpha_{\tau}}+\frac{\eta_T-\eta_s}{\beta_{p}}=\frac{\alpha_\tau T}{c_{p,\tau}},
\end{equation}
and, for $dp=0$,
\begin{equation}
\eta_T=\eta_s+\frac{\alpha_\tau \beta_{p} T}{c_{p,\tau}}
\end{equation}
and
\begin{equation}
\frac{k_{T,\gamma}}{k_{s,\tau}}=\frac{\alpha_{\gamma}}{\alpha_{\tau}}.
\end{equation}

\subsubsection{\boldmath$d\epsilon(T,p,\gamma)=d\epsilon(s,p,\gamma)$}
From this equality, we obtain
\begin{equation}
\label{eq:voltransf3}
\frac{k_{T,\gamma}-k_{s,\gamma}}{\alpha_{\gamma}}+\frac{1}{\beta_p}\left(\eta_{T}-\eta_s\frac{S_{T,p}}{S_{s,p}}\right)=\frac{\alpha_{\gamma} T}{c_{p,\gamma}}.
\end{equation}
For $dp=0$, this simplifies to
\begin{equation}
\eta_{T}=\eta_s\frac{S_{T,p}}{S_{s,p}}+\frac{\alpha_{\gamma} \beta_p T}{c_{p,\gamma}},
\end{equation}
and
\begin{equation}
k_{T,\gamma}=k_{s,\gamma}.
\end{equation}

For $d\gamma=0$, (\ref{eq:voltransf3}) reduces to
\begin{equation}
\label{eq:krel2}
k_{T,\gamma}=k_{s,\gamma}+\frac{\alpha_{\gamma}^2 T}{c_{p,\gamma}},
\end{equation}
and
\begin{equation}
\frac{\eta_{T}}{\eta_s}=\frac{S_{T,p}}{S_{s,p}}.
\end{equation}
Similarly to (\ref{eq:krel1}), equation (\ref{eq:krel2}) is here shown to be valid also for a generic state of constant plane shear strain. 

%From $d\epsilon(T,p,\tau)=d\epsilon(s,p,\gamma)$,
%\begin{equation}
%\frac{1}{\beta_{p}}\left(\eta_T-\eta_s\frac{S_{T,p}}{S_{s,p}}\right)=\frac{\alpha_{\gamma} T}{c_{p,\gamma}},
%\end{equation}
%and along an isobar, $dp=0$,
%\begin{equation}
%\frac{k_{t,\tau}}{\alpha_{\tau}}=\frac{k_{s,\gamma}}{\alpha_{\gamma}}.
%\end{equation}
\subsubsection{\boldmath$d\epsilon(T,p,\tau)=d\epsilon(T,p,\gamma)$}
This condition leads to
\begin{equation}
\label{eq:voltransf4}
\frac{k_{T,\tau}}{\alpha_{\tau}}-\frac{k_{s,\gamma}}{\alpha_{\gamma}}+\frac{1}{\beta_{p}}\left(\eta_{T}-\eta_s\frac{S_{T,p}}{S_{s,p}}\right)=\frac{\alpha_\gamma T}{c_{p,\gamma}},
\end{equation}
which along an isobar, $dp=0$, reduces to
\begin{equation}
\eta_{T}=\eta_s\frac{S_{T,p}}{S_{s,p}}+\frac{\alpha_\gamma \beta_{p} T}{c_{p,\gamma}},
\end{equation}
 and therefore
 \begin{equation}
\frac{k_{T,\tau}}{k_{s,\gamma}}=\frac{\alpha_{\tau}}{\alpha_{\gamma}}.
 \end{equation}
 
\subsection{Shear-angle representation}
In the shear-angle representation, the total differential of $\gamma$ can be expressed as a function of a choice of independent variables in terms of various material property coefficients as
\begin{eqnarray}
d\gamma=\beta_{p}dT-\eta_Tdp+S_{T,p}d\tau\\ 
d\gamma=\frac{\beta_p T}{c_{p,\tau}}ds-\eta_s dp+S_{s,p}d\tau \\
d\gamma=\beta_\epsilon dT +\frac{\eta_T}{k_{T,\tau}}d\epsilon +S_{T,\epsilon}d\tau \\ 
d\gamma= \frac{\beta_\epsilon T}{c_{v,\tau}}ds+\frac{\eta_s}{k_{s,\tau}}d\epsilon+S_{s,\epsilon}d\tau.
\end{eqnarray}
Expressions relating shear properties to thermal and pressure properties are obtained by comparing the above expressions.

\subsubsection{\boldmath$d\gamma(T,p,\tau)=d\gamma(s,p,\tau)$}
From this equation, diving by $ds$ at constant $T$, we obtain
\begin{equation}
\label{eq:angltransf1}
\frac{S_{T,p}-S_{s,p}}{\beta_p}+\frac{\eta_T-\eta_s}{\alpha_{\tau}}=\frac{\beta_p T}{c_{p,\tau}}.
\end{equation}
At constant $\tau$, the relation yields again
\begin{equation}
\eta_T=\eta_s+\frac{\alpha_{\tau} \beta_p T}{c_{p,\tau}},
\end{equation}
and hence the equivalence of isothermal and adiabatic, isobaric shear compliances,
\begin{equation}
\label{eq:eqmoduli1}
S_{T,p}=S_{s,p}.
\end{equation}
While previous work has considered relationship (\ref{eq:eqmoduli1}) to hold in any conditions \cite{landau1965theory,poirier2000introduction,dyre2006colloquium}, the above derivation shows that it is valid only in isoshear-stress transformations.
At constant $p$, (\ref{eq:angltransf1}) reduces to
\begin{equation}
\label{eq:burnsp}
S_{T,p}=S_{s,p}+\frac{\beta_p^2 T}{c_{p,\tau}},
\end{equation}
and thus
\begin{equation}
\eta_T=\eta_s.
\end{equation}

\subsubsection{\boldmath$d\gamma(s,p,\tau)=d\gamma(T,\epsilon,\tau)$}
Similarly, one has
\begin{equation}
\label{eq:angltransf2}
\frac{S_{T,\epsilon}}{\beta_\epsilon}-\frac{S_{s,p}}{\beta_p}+\frac{\eta_T-\eta_s}{\alpha_\tau}=\frac{\beta_p T}{c_{p,\tau}}.
\end{equation}
For constant $\tau$, this simplifies to
\begin{equation}
\eta_T=\eta_s+\frac{\alpha_\tau \beta_p T}{c_{p,\tau}},
\end{equation}
and thus
\begin{equation}
\label{eq:Sbeta1}
\frac{S_{s,p}}{S_{T,\epsilon}}=\frac{\beta_p}{\beta_\epsilon}.
\end{equation}

\subsubsection{\boldmath$d\gamma(T,\epsilon,\tau)=d\gamma(s,\epsilon,\tau)$}
Furthermore, in this case we have
\begin{equation}
\label{eq:angltransf3}
\frac{S_{T,\epsilon}-S_{s,\epsilon}}{\beta_{\epsilon}}+\frac{1}{\alpha_{\tau}}\left(\eta_T-\eta_s\frac{k_{T,\tau}}{k_{s,\tau}}\right)=\frac{\beta_{\epsilon} T}{c_{v,\tau}}.
\end{equation}
At constant $\epsilon$, it simplifies to \cite{burns2018elastic}
\begin{equation}
\label{eq:burnsv}
S_{T,\epsilon}=S_{s,\epsilon}+\frac{\beta_{\epsilon}^2 T}{c_{v,\tau}},
\end{equation}
which was obtained in implicit conditions of constant volume. From equation (\ref{eq:angltransf3}) and $d\epsilon=0$ one also has
\begin{equation}
\frac{\eta_T}{\eta_s}=\frac{k_{T,\tau}}{k_{s,\tau}}.
\end{equation}
At constant $\tau$,
\begin{equation}
\eta_T=\eta_s\frac{k_{T,\tau}}{k_{s,\tau}}+\frac{\alpha_{\tau} \beta_{\epsilon} T}{c_{v,\tau}},
\end{equation}
hence the equivalence of isothermal and adiabatic, isochoric compliances,
\begin{equation}
\label{eq:eqmoduli2}
S_{T,\epsilon}=S_{s,\epsilon}.
\end{equation}
As for (\ref{eq:eqmoduli1}), we emphasize again that equation (\ref{eq:eqmoduli2}) is obtained strictly for isoshear-stress transformations.

\subsubsection{\boldmath$d\gamma(T,p,\tau)=d\gamma(s,\epsilon,\tau)$}
As before, from this equality one also has
\begin{equation}
\label{eq:Sbeta2}
\frac{S_{T,p}}{\beta_p}-\frac{S_{s,\epsilon}}{\beta_\epsilon}+\frac{1}{\alpha_{\tau}}\left(\eta_T-\eta_s\frac{k_{T,\tau}}{k_{s,\tau}}\right)=\frac{\beta_{\epsilon} T}{c_{v,\tau}},
\end{equation}
which simplifies along an isoshear, $d\tau=0$, as
\begin{equation}
\eta_T=\eta_s\frac{k_{T,\tau}}{k_{s,\tau}}+\frac{\alpha_{\tau} \beta_{\epsilon} T}{c_{v,\tau}},
\end{equation}
and thus
\begin{equation}
\frac{S_{T,p}}{S_{s,\epsilon}}=\frac{\beta_p}{\beta_\epsilon}.
\end{equation}

The general relations above, (\ref{eq:angltransf1}), (\ref{eq:angltransf2}), (\ref{eq:angltransf3}) and (\ref{eq:Sbeta2}), extend the one presented by Burns \cite{burns2018elastic} between isothermal and adiabatic shear compliances at constant $\epsilon$. In fact, they also include the isothermal and adiabatic coefficients of dilatancy, $\eta_{T}$ and $\eta_s$, and are derived for either isobaric or isochoric transformations.

\section{Conclusions}

We have provided 12 general relationships among the 18 properties of materials (6 of which are independent) that exist in conditions of plane shear. The other shear angles remain constant, so that during a thermodynamic transformation the shear stresses in those directions do not do any work. We discuss infinitesimal, reversible transformations around a generic state, described by the fundamental equation $u=u(s,\epsilon,\gamma)$ or $g=g(T,p,\tau)$, and extend previous work \cite{burns2018elastic} by introducing the Reynolds dilatancy coefficient, $\eta$, which expresses the pressure-shear stress coupling \cite{reynolds1885lvii}. This gives rise to new general relations between thermal, pressure, and shear material properties. Their importance ranges from analysis of nano-materials behavior to characterization of earthquake precursors \cite{nur1972dilatancy,barnes1989shear}
%The actual presence and possible effects of this coupling will be explored in future work. 

For $\eta=0$, we recover the relationship between isothermal and adiabatic shear compliances and the coefficient of thermal shear deformation \cite{burns2018elastic}, see (\ref{eq:burnsv}), and show that this relation is generally valid at both constant pressure or constant logarithmic strain, see equations (\ref{eq:burnsp}) and (\ref{eq:burnsv}). New relationships between isobaric and isochoric shear thermal deformation and the compliances are derived, e.g., equations (\ref{eq:angltransf2}) and (\ref{eq:Sbeta2}). Our results also extend the well-known relation between the isothermal and adiabatic compressibility to a generic state of plane shear, see (\ref{eq:krel1}) and (\ref{eq:krel2}).

Imposing $\tau=0$ and $\gamma=0$ in the derived general relations, one returns to the thermodynamics in hydrostatic conditions, where both $\eta$ and $\beta$ are considered to be zero. 
%where isothermal and adiabatic shear compliances (and moduli) are equal \cite{burns2018elastic,landau1965theory} and iso-$\tau$ or iso-$\gamma$ transformations become equivalent, see for example (\ref{eq:burnsp}) and (\ref{eq:burnsv}). 
The material properties in fact reduce to the two heat capacities (constant volume or pressure), a unique coefficient of thermal expansion, two compressibilities (isothermal and adiabatic) and a shear compliance. Their relations however remain formally the same also outside of the pure hydrostatic conditions, e.g., (\ref{eq:krel1}), although their numerical values depend upon the specific constant values of $\tau$ or $\gamma$ at which the material is maintained. The neighborhood of this stress state is the elastic regime, in which $k_{T,\tau}$ and $S_{p,\tau}$ are assumed constant.

Future work will deal with embedding the obtained relationships within a continuum-mechanics representation with local equilibrium assumptions and extend them to general non-homogeneous configurations. We hope that our analysis might help shed light on phase transitions in the presence of shear and towards a thermodynamic representation of the glass transition.

\section{Acknowledgments}
This work was supported through the National Science Foundation (NSF) grants EAR-1331846 and FESD-1338694.
	% put your acknowledgments here.

%\bibliographystyle{apalike}
%\bibliography{RefthermoSalvo}

\end{document}